\documentclass[fleqn,usenatbib]{mnras}

\usepackage{newtxtext,newtxmath}

\usepackage[T1]{fontenc}
\usepackage{booktabs}

\usepackage{amssymb}
\usepackage{tabularx, siunitx}
\usepackage[flushleft]{threeparttable}
\usepackage[normalem]{ulem}
\usepackage{enumitem}

\DeclareRobustCommand{\VAN}[3]{#2}
\let\VANthebibliography\thebibliography
\def\thebibliography{\DeclareRobustCommand{\VAN}[3]{##3}\VANthebibliography}

\usepackage{graphicx}	
\usepackage{amsmath}	
\usepackage{multirow}   
\usepackage{newtxmath}
\usepackage{rotating}
\usepackage{pdflscape}

\title[Isolating Broadband Radio Technosignatures (BRaTs)]{Isolating Broadband Radio Technosignatures (BRaTs): A Framework for Detecting Planetary-Scale Leakage}

\author[M.A. Garrett et al.]{Michael A. Garrett$^{1,2,3}$ 
\\
$^{1}$Jodrell Bank Centre for Astrophysics, Department of Physics and Astronomy, Alan Turing Building, University of Manchester, Oxford Road, M13 9PL, UK\\
$^{2}$Leiden Observatory, Leiden University, P.O. Box 9513, NL-2300 RA Leiden, The Netherlands\\
$^{3}$University of Malta, Institute of Space Sciences and Astronomy, Msida, MSD2080, Malta\\
}

\pubyear{2026}

\begin{document}

\label{firstpage}
\pagerange{\pageref{firstpage}--\pageref{lastpage}}
\maketitle


\begin{abstract}

The search for extraterrestrial intelligence (SETI) has traditionally focused on the detection of narrowband electromagnetic beacons. However, terrestrial technology is increasingly evolving toward distributed, low-power, wideband digital infrastructure. The strict adherence to narrowband filtering that characterises most SETI surveys, therefore, risks discarding the aggregate leakage signatures of advanced civilisations by systematically misclassifying them as unstructured noise. We investigate the feasibility of detecting such planetary-scale broadband radio technosignatures (BRaTs) using a hierarchical observational framework. In this tiered approach, wide-field radio surveys conducted by next-generation arrays (such as the SKA and its precursors) perform the initial deep-field observations, with targeted Very Long Baseline Interferometry (VLBI) providing the definitive, high-resolution follow-up. Because broadband continuum emission is largely insensitive to Doppler drift, long-duration ``SETI Deep Fields'' are observationally viable, extending the accessible detection volume for Kardashev Type~I leakage to $\sim 100$~pc. To distinguish these signals from other astrophysical confounders, a multi-parameter diagnostic framework is proposed. Candidate technosignatures are identified through a convergence of high brightness temperatures, negligible circular polarisation, spectral non-uniformity, interstellar scintillation, and sub-milliarcsecond astrometric co-motion with nearby Galactic stars/exoplanets. 

\end{abstract}




\section{Introduction}

Since its inception, the Search for Extraterrestrial Intelligence (SETI) has traditionally focused on detecting narrowband radio and optical signals \citep{1961PhT....14d..40D, 1961Natur.190..205S, 1973Icar...19..329V}. These are characterised by their very narrow spectral occupancy, typically assumed to be only a few Hertz wide. The major advantage of searching for such signals is that they are easily distinguishable from the natural, broadband cosmic radio background. The narrowest features associated with cosmic radio sources are Galactic masers, even these show structure only on scales $> 500$~Hz \citep{1987MNRAS.225..491C}. Thus, most modern radio and optical SETI surveys continue to focus on the detection of narrowband signals, often targeting local stars \citep{enriquez_breakthrough_2017, 2017AJ....153..251T, price_breakthrough_2020}, the galactic centre ~\citep{2021AJ....162...33G}, and nearby galaxies \citep{garrett_constraints_2023, choza_breakthrough_2024}. The goal is specific - to detect deliberately transmitted, high-power, narrowband beacons.

Over the past 50 years, radio leakage from our own technological civilisation has changed markedly, evolving from a relatively small number of powerful narrowband transmitters, such as national radio and television broadcasts \citep{1978Sci...199..377S}, to a vast and heterogeneous population of much lower-power broadband emitters. These include cellular base stations, mobile devices, and WiFi systems \citep{2023MNRAS.522.2393S}, a trend that is expected to continue with the development of future communication standards such as 6G. In addition, civilian and military radar and navigation systems, satellite constellations, and other broadband technologies now contribute to a highly complex and overlapping electromagnetic environment \citep{2026AAS...24733103F, Saide2026}. The methods for encoding information in communication signals have also evolved significantly over the last few decades. Familiar analogue modulation techniques, such as amplitude modulation (AM) and frequency modulation (FM), have given way to more complex digital modulation schemes \citep{haykin2001communication}.

If advanced extraterrestrial civilisations deploy broadband communication and navigation systems of comparable or greater scale than those on Earth, their aggregate unintentional radio leakage could produce detectable emission spanning gigahertz-scale bandwidths. Under plausible assumptions, such broadband signals may remain detectable across interstellar distances \citep{2023MNRAS.522.2393S}. The potential relevance of wide-band radio emission to SETI was first noted by \citet{1980JBIS...33..391C}. Building on this, Messerschmitt and Morrison developed a theoretical framework for detecting extraterrestrial broadband transmissions using matched filtering techniques \citep{2012AcAau..78...80M, 2012AcAau..78...90M}. In matched filtering, receiver sensitivity increases as prior knowledge of the transmitted waveform improves. In the ideal case the receiver knows the signal structure exactly, as occurs in terrestrial digital communication systems where the modulation and coding schemes are predefined. In SETI searches, however, the waveform of a putative extraterrestrial transmission is unknown. Either a simplified assumption must therefore be adopted, such as the monochromatic tone implicitly assumed in narrowband SETI, or the waveform must be recovered through an extensive search over possible signal structures. Given that an advanced civilisation may employ numerous communication technologies, some of which may be unfamiliar to us (see \citet{2026AcAau.238..160G}), the computational cost of exploring this parameter space is currently prohibitive for large-scale searches.

In recent years, observational efforts have begun to expand the SETI parameter space beyond traditional narrowband searches to include pulsed and periodic broadband radio signals \citep{2022ApJ...932...81G, 2023AJ....165..255S}. Earlier theoretical studies of broadband radio leakage and its detectability at low frequencies \citep{2007JCAP...01..020L} also helped motivate technosignature surveys with the Murchison Widefield Array (MWA), including searches conducted with moderate spectral resolution ($\sim$10 kHz). More recently, \citet{2026ApJ..1001..167Z} discussed the detection of deliberate wideband transmissions. 

The framework considered here differs from these approaches by focusing on the identification of faint, compact broadband continuum sources through deep interferometric imaging, targeted VLBI follow-up, and multi-wavelength source characterisation.  Rather than attempting to detect individual structured transmissions, we consider the possibility that the aggregate output of many independent technological systems may produce detectable broadband radio emission. The superposition of emissions from millions of incoherent emitters, e.g., communication links, navigation systems, radar transmissions, and other technological infrastructure distributed across a planetary surface and low-Earth orbit, would generate a spectrally extended signal that appears observationally as faint continuum radio emission, easily mistaken for unstructured noise by conventional SETI searches. Within this framework, a technologically active planet would appear not as an isolated, discrete transmission, but rather as a low-intensity, artificially generated radio continuum source, most likely spatially coincident with a star/planetary system located within our own Galaxy.

In this paper, we investigate the scientific basis for detecting  broadband radio technosignatures (hereafter BRaTs). In contrast to traditional SETI searches, we consider the regime in which the aggregate emission from many independent technological systems appears observationally as faint, broadband radio continuum emission associated with a planetary system. A central question addressed in this work is how such emission could be distinguished from the population of faint natural radio sources that dominate the centimetre-wavelength source counts at sub-mJy and $\mu$Jy flux densities. 

Section~2 estimates the expected broadband radio leakage from Kardashev Type~I civilisations and evaluates its detectability by radio interferometers, including the Square Kilometre Array (SKA) and global Very Long Baseline Interferometry (VLBI) networks. We introduce the concept of SETI deep fields, and also discuss how milliarcsecond angular resolution combined with sub-milliarcsecond astrometry permits the nature and precise localisation of technological aggregates to be determined. Section~3 examines the properties of BRaTs and how these can be distinguished from the principal astrophysical confounders (low-luminosity active galactic nuclei, distant star-forming galaxies, radio-loud stars, magnetised exoplanets, pulsars, etc.). Section 4 presents a multi-parameter discrimination framework based on spectral behaviour, brightness temperature, polarisation properties, temporal variability, scintillation, and propagation effects. We also discuss extending our methodology to Kardashev Type II BRaTs, the effects of the ionospheric cut-off for exoplanets hosted by M-dwarf stars, and AI anomaly detection. 
Finally, Section~5 summarises the main conclusions and discusses the implications of broadband radio surveys for the future evolution of technosignature search strategies.

\section{The detectability of Kardashev Type I civilisations as broadband technosignatures}
\label{section2}

\subsection{Estimating the broadband radio leakage from a Kardashev Type I civilisation} 
\label{K1}

The present-day global energy consumption of human civilisation is estimated at $\sim 6 \times 10^{20}$~J~yr$^{-1}$ \citep{EI2024}. This is equivalent to an average power of $P_{\mathrm{tot}} \sim 2 \times 10^{13}$~W, placing the Earth roughly as a Kardashev Type~0.7 civilisation \citep{1973Icar...19..350S}. Simulations of terrestrial radio leakage from mobile communication systems and civilian/military radars \citep{2023MNRAS.522.2393S, Saide2026} suggest that the Earth currently produces an aggregate spectral luminosity of the order $L_\nu \sim 400$~W~Hz$^{-1}$ at L-band. If we assume broadband leakage scales linearly with total power consumption, we estimate a radio leakage luminosity of $L_\nu \sim 4\times10^{5}$~W~Hz$^{-1}$. 

Consider an extraterrestrial civilisation located at distance $D$ whose technological infrastructure produces isotropic-equivalent broadband leakage with spectral luminosity $L_\nu$ across a bandwidth $\Delta\nu$. The flux density measured here on  Earth is then:

\begin{equation}
S = \frac{L_\nu}{4\pi D^2}
\end{equation}

For a source with $L_\nu \sim 4\times10^{5}$~W~Hz$^{-1}$, located at $D = 10$~pc, the resulting aggregate flux density is approximately $33~\mu$Jy. By $D = 60$~pc the source is $\sim 1~\mu$Jy. While these represent extremely faint emission levels, modern deep-field radio surveys are already probing the sub$\mu$Jy regime , and next generation arrays will surpass this \citep{2024ApJ...972...89J}.


\subsubsection{Sensitivity of VLBI to BRaTs}

An artificial broadband signal originating from an Earth-like exoplanet would appear as a spatially compact point source. In a VLBI array such a signal would be detectable across global baselines and would therefore be observed as an unresolved emitter with a high inferred brightness temperature ($T_b$). The corresponding brightness temperature measured with a VLBI array with synthesised beam solid angle $\Omega$ is:

\begin{equation}
T_b = \frac{S c^2}{2k_B \nu^2 \Omega}.
\end{equation}

For a Gaussian synthesised beam,

\begin{equation}
\Omega = 1.133\,\theta_B^2
\end{equation}

where $\theta_B$ is the FWHM beam size. Since $\theta_B \approx \lambda/B$ and $\lambda=c/\nu$, a lower limit on the brightness temperature of an unresolved source can be written as:

\begin{equation}
T_b > \frac{S B^2}{2.266\,k_B}.
\end{equation}

Global VLBI arrays routinely achieve milliarcsecond resolution, and microJy sensitivities over wide fields of view \citep{2023MNRAS.519.1732N, 2025arXiv250419579R}. Using the EVN sensitivity calculator \citep{EVNcalculator}, a global VLBI array observing at C-band ($\sim4$--$8$~GHz) achieves a $1-\sigma$ r.m.s. sensitivity of $\sim 1~\mu$Jy assuming a recording data rate of 2 Gbps and an on-source observing time of 10 hours. A Kardashev Type 1 civilisation, presenting leakage radiation at a level of $\sim 34\mu\mathrm{Jy} $ (see section~\ref{K1}), is therefore well detected with this set-up. Since any leakage is expected to be unresolved at a distance of 10~pc, we can only determine a lower limit on its brightness temperature,   $T_b > 1.1 \times 10^6 \,\mathrm{K} $. If the emission originates from a region comparable in size to an Earth-like planet (diameter $\sim1.2\times10^4$ km), the intrinsic brightness temperature would be substantially higher, of order $\sim10^{10}\,\mathrm{K}$.

We note that the sensitivity of global VLBI arrays is expected to improve significantly in the near future with the inclusion of SKA precursors (e.g. MeerKAT) and the initial deployment of the SKA itself \citep{2019arXiv190110361P, 2019arXiv190308627G}. Sensitivity estimates based on the EVN Sensitivity Calculator \citep{EVNcalculator} indicate that incorporating the SKA baseline design (SKA-AA4) will approximately double the array sensitivity, yielding a $1\sigma$ r.m.s. noise level of $\sim 0.5~\mu$Jy.

\subsubsection{SETI Deep Fields}

Traditional narrowband technosignature searches are fundamentally complicated by Doppler drift - this arises from the relative motion between the transmitter and the observer \citep{2019ApJ...884...14S}. Since the value of the drift rate is unknown \textit{a priori}, SETI analysis pipelines are forced to execute computationally expensive acceleration trials to mitigate signal smearing across narrow frequency channels and preserve detection sensitivity. In contrast, observations of BRaTs are largely insensitive to this effect, as by definition, they target an aggregate of emission across a wide range of frequency space. This insensitivity eliminates the need for the extremely fine spectral channelisation and computationally intensive acceleration searches that characterise conventional SETI experiments. More importantly, it permits exceptionally long coherent integrations. This unique advantage enables us to introduce the concept of "SETI Deep Fields"—prolonged observational campaigns analogous to those used to detect, for example, galaxy formation in the early universe, e.g., \citep{1996AJ....112.1335W}.

In principle, integration times spanning many days are already feasible using established wide-field techniques \citep{2001A&A...366L...5G, 2005ApJ...619..105G, 2013A&A...550A..68C, 2013A&A...551A..97M, 2018A&A...616A.128H, 2019A&A...625C...1R, 2024MNRAS.529.2428D} combined with advanced multi-source calibration \citep{2016A&A...587A..85R}. Crucially, these calibration methods can extend coherence times indefinitely, permitting arbitrarily long on-sky integrations. If total integration times of the order of $\sim1000$ hours are achieved through deep commensal surveys, sensitivities will improve by a factor of 10. When combined with the full capabilities of SKA AA4, such deep integrations would extend the detectable range of Kardashev Type~I leakage to distances of up to $ 100$~pc with detections being made at a signal-to-noise ratio of $\sim 7$. We note that the ngVLA can achieve similar sensitivities. Naturally, these prolonged observational campaigns would simultaneously facilitate a wealth of ancillary high-resolution radio data. 


\subsubsection{Signal localisation and proper motion}

Phase-referenced VLBI observations provide astrometric precision at the sub-milliarcsecond level, enabling accurate localisation of compact broadband emitters. A good example of this has been the localisation of Fast Radio Bursts (FRBs) using VLBI techniques \citep{2020Natur.577..190M}. At a distance of $100$~pc, an angular scale of $1$~mas corresponds to a linear distance of approximately $0.1$~AU. Such precision allows candidate technosignatures to be associated directly with specific components of planetary systems.

For nearby stellar systems ($10$--$100$~pc), sub-milliarcsecond astrometry corresponds to linear scales of $\sim0.001$--$0.01$~AU. This resolution is sufficient to distinguish emission originating from the stellar surface (e.g.\ stellar flares) from emission associated with a planetary orbit. Such localisation therefore provides an important diagnostic for identifying planetary-scale technological emitters.

VLBI astrometry can also detect orbital motion in compact radio sources via multi-epoch observations made on timescales of months to years \citep{2018arXiv181007235G}. For example, a planet with an orbital radius of $0.01$~AU at a distance of $100$~pc subtends an angular displacement of $\sim0.1$~mas. Tracking such motion offers an additional method for confirming the association of a radio source with an exoplanetary system. We note that similar techniques may also allow the detection and tracking of artificial spacecraft emitting powerful radio signals, with observable motion occurring on timescales ranging from years for slow-moving objects to days for relativistic trajectories.

\subsubsection{Cross-matching with astronomical catalogues}

The precise localisation afforded by VLBI enables robust cross-identification with existing astronomical catalogues. It will be important that any SETI Deep Field projects target areas of the sky with excellent auxiliary multi-waveband coverage. Candidate broadband emitters can be compared with stellar catalogues such as \emph{Gaia} \citep{gaia_collaboration_gaia_2023} to determine whether their milliarcsecond-accurate positions coincide with known stars or offset planetary systems. We will return to this discussion in the next section.

\section{Discrete Astrophysical Confounders}
\label{confounders}

The identification of BRaTs in the 1--10~GHz regime requires discrimination against natural compact radio emitters. Although several astrophysical processes can generate high brightness temperatures and broadband emission that masquerades as unstructured noise, they differ systematically from a planetary-scale technological source in their spectral behaviour, polarisation properties, propagation signatures, and temporal variability. Often, no single observable is decisive; rather, a consistent set of diagnostics must be considered when evaluating candidate signals. 

In this section, we evaluate the principal astrophysical confounders present in deep radio surveys and establish the key discriminating features required to distinguish them from genuine technosignatures. We first of all list the likely properties of BRaTs. 

\subsection{Properties of BRaTs}

We expect genuine broadband radio technosignatures (BRaTs) to exhibit the following defining characteristics:  

\begin{enumerate}[label=(\roman*), leftmargin=*]
    \item High brightness temperature and compact morphology: As outlined in Section~\ref{section2}, planetary-scale BRaTs will be observed by global VLBI networks as faint, unresolved point sources, yielding inferred brightness temperatures of $T_b \gtrsim 10^6$~K. 
    
    \item Galactic kinematics: A local technosignature ($10$--$100$~pc) will probably be gravitationally bound to a nearby stellar system. Consequently, it will exhibit significant proper motion and annual parallax ($\gtrsim 10$~mas~yr$^{-1}$), alongside potential orbital kinematics.  
    
    \item Spectral shape: Because technological aggregates are shaped by engineering constraints rather than astrophysics, their broadband emission is expected to display distinct spectral non-uniformity. This includes ``notches'' corresponding to planetary atmospheric absorption, sharp cut-offs dictated by regulated frequency allocations, or localised regions of enhanced power resulting from specific high-energy applications (e.g., microwave power transmission).
    
    \item Polarisation: Negligible circular or linear polarisation, reflecting the incoherent aggregation of millions of independent, randomly oriented transmitters.
    
    \item Interstellar scintillation (ISS) and temporal modulation: A planetary-scale emitter at a distance of $10$~pc subtends an angular size of only $\sim 8~\mu$as, acting as an extreme point source. This will subject the signal to strong, broadband ISS \citep{1990ARA&A..28..561R}, producing stochastic amplitude variability as predicted by \citet{1997ApJ...487..782C} and \citet{2023ApJ...952...46B}. Crucially, these short-term propagation effects will likely be superimposed upon a highly stable, long-term diurnal modulation associated with a rotating host planet.  
    \item Host system identification: Through precise VLBI astrometry, candidate BRaTs can be unambiguously cross-matched with nearby Galactic stars using optical and infrared catalogues—most notably \emph{Gaia} \citep{gaia_collaboration_gaia_2023}. This precise localisation will  confirm their origin within the local Milky Way. Small offsets from \emph{Gaia} stellar positions are expected. 
\end{enumerate}

\subsection{Extragalactic radio sources} 

The dominant confounding sources to be found in deep centimetre-wave radio surveys is extragalactic in origin, primarily comprising Starforming galaxies (SFGs) and Active Galactic Nuclei (AGN). We consider here how their characteristics compare to BRaTs.

\subsubsection{Starforming galaxies (SFGs)}

Starforming galaxies dominate the faint radio source population at flux densities $S_{\nu} \lesssim 100~\mu$Jy \citep{2025A&A...697A..81G}. Their radio emission is primarily due to non-thermal synchrotron radiation from cosmic-ray electrons accelerated in supernova remnants. Despite their abundance in deep radio surveys, SFGs can be readily distinguished from BRaTs through several key observational properties:

\begin{enumerate}[label=(\roman*), leftmargin=*]

\item Brightness temperature: SFGs are generally resolved by VLBI observations and exhibit relatively low brightness temperatures, typically $T_b \lesssim 10^5$--$10^6$~K \citep{1992ARA&A..30..575C}.

\item Infrared--radio correlation: SFGs follow the well-established infrared--radio correlation in both the local \citep{1973A&A....29..263V, 1993ApJ...415...93H} and high-redshift universe \citep{2002A&A...384L..19G, 2004ApJS..154..147A, 2009ApJ...706..482M, 2010A&A...518L..31I}, and exhibit extended optical morphologies consistent with star-forming galaxies.

\item Radio morphology: SFGs typically have angular sizes of $\sim 0.5$--$2''$, with radio emission tracing the optical disc of the host galaxy \citep{2005MNRAS.358.1159M}.

\item Cosmological distances: Multi-wavelength cross-matching with wide-area surveys (e.g.\ \citet{2014yCat.2328....0C, 2016MNRAS.460.1270D}) will associate many candidate BRaTs with extended galaxies exhibiting natural spectral energy distributions (SEDs). Subsequent photometric and spectroscopic redshift measurements will confirm that these sources reside at extragalactic distances, consistent with the background population of faint radio galaxies.

\end{enumerate}

\subsubsection{Active Galactic Nuclei (AGN)}

While SFGs are more numerous, AGN also represent a significant confounder and are well known for their small angular sizes \citep{1971MNRAS.152..477M}.  Below $100~\mu$Jy, Radio-Quiet (RQ) AGN begin to outnumber Radio-Loud (RL) systems \citep{2017ApJ...842...95M}. These sources often host weak, compact active nuclei with $T_b > 10^6$~K, and they can remain unresolved even on global VLBI baselines. 
Despite this, AGN can be disqualified as technosignatures through several key filters:

\begin{enumerate}[label=(\roman*), leftmargin=*]

\item Astrometric kinematics: AGN are located at cosmological distances and are effectively stationary on the sky, exhibiting no measurable parallax or proper motion \citep{2016AJ....152..118H}. 

\item Spectral behaviour: AGN typically exhibit smooth, power-law spectra ($S_\nu \propto \nu^\alpha$) governed by synchrotron emission processes \citep{1973raas.book.....K}. This behaviour contrasts strongly with the expected spectral complexity of BRaTs.

\item Interstellar scintillation (ISS): A subset of extremely compact AGN cores exhibit ISS \citep{2002Natur.415...57D}. However, for the majority of sources, intrinsic source sizes ($\gtrsim 10^{15}$~cm) suppress or ``quench'' scintillation-induced variability.

\item Cosmological distances: As with SFGs, AGN will also be identified with distant host galaxies using wide-area multi-wavelength surveys. Despite their compact radio morphology, optical and infrared data will exhibit the characteristic spectral energy distributions of accreting supermassive black holes and permit redshift determinations that securely place these sources at extragalactic distances.

\end{enumerate}

\subsection{Exoplanetary and Stellar Radio Emission}

While extragalactic sources are by far the most numerous confounders, local Galactic sources—specifically magnetised exoplanets and active stars present a different challenge. These sources can exhibit high brightness temperatures and are astrometrically "local," sharing the proper motion of the target system. However, they are physically constrained by plasma physics and magnetic field strengths in ways that technological leakage is not.

\subsubsection{Magnetised Exoplanets}

Exoplanetary radio emission is generated via the electron--cyclotron maser instability (CMI), the mechanism responsible for the auroral radio emission of Solar System planets \citep{1998JGR...10320159Z, 2007P&SS...55..598Z}. However, these natural emissions possess distinct physical characteristics that allow them to be definitively distinguished from broadband radio technosignatures (BRaTs).

\begin{enumerate}[label=(\roman*), leftmargin=*]

    \item Frequency cut-offs: CMI emission is strictly bounded by the local electron cyclotron frequency, $\nu_{\mathrm{ce}}[\mathrm{MHz}] \approx 2.8\, B[\mathrm{G}]$ \citep{1985ARA&A..23..169D}. For this natural emission to contaminate the $1$--$10$~GHz window, an exoplanet would require extreme magnetic field strengths of $B \gtrsim 350$--$3500$~G. Such values are orders of magnitude higher than Jupiter's polar field ($\sim 14$~G) and far exceed theoretical estimates of the maximum plausible exoplanetary field limit of $\sim 100$~G \citep{2021ApJ...908...77H}.

    \item Temporal and spectral behaviour: Planetary CMI is fundamentally sporadic, characterised by transient bursts with rapidly drifting dynamic spectra (e.g., Jovian decametric arcs) \citep{2023NatCo..14.5981M}. This erratic natural emission is entirely distinct from the persistent and stable broadband behaviour defining candidate BRaTs.
    
    \item Polarisation: CMI emission is intrinsically highly circularly polarised (often approaching $100\%$) in contrast to the aggregate leakage of BRaTS which we expect to be unpolarised.
    
\end{enumerate}

\subsubsection{Active Stars and M-Dwarfs}

Late-type stars, particularly M-dwarfs and flare stars, are prolific radio emitters across the GHz regime, producing both incoherent gyrosynchrotron radiation and coherent bursts \citep{1990ApJ...353..265B, 2019ApJ...871..214V}. They can be distinguished from technosignatures through the following:

\begin{enumerate}[label=(\roman*), leftmargin=*]

    \item Astrometric Offset: At distances of $10$--$100$~pc, a planet in an AU-scale orbit exhibits an angular offset of $\sim 10$--$100$~mas from its host star. VLBI’s sub-milliarcsecond precision can easily distinguish between emission originating from the stellar corona and emission associated with a planetary companion.
    \item Spectral Shape: Stellar gyrosynchrotron emission typically displays a smooth, peaked spectrum (transitioning from optically thick to thin).
    \item Variability: Stellar radio activity is largely stochastic and characterized by high-amplitude flares. 
\end{enumerate}

\subsubsection{Ultracool Dwarfs (Brown Dwarfs)}

Ultracool dwarfs (UCDs), comprising late M, L, and T spectral types, bridge the physical gap between low-mass stars and massive exoplanets. Unlike Solar System planets, UCDs frequently possess surface magnetic fields of several kilogauss \citep{2012ApJ...746...23M}. Consequently, they are potent sources of CMI emission that routinely extends well into the $4$--$10$~GHz regime, directly overlapping with the optimal terrestrial microwave window for SETI. 

They can be distinguished from technological leakage through the following observables:
\begin{enumerate}[label=(\roman*), leftmargin=*]

    \item Polarisation: The CMI emission from UCDs is intrinsically highly circularly polarised, frequently approaching 100\% \citep{2001Natur.410..338B}, which starkly contrasts with the incoherent, unpolarised nature of aggregate technological leakage.
    \item Temporal Behaviour: UCD radio emission is characterised by strict rotational modulation, typically manifesting as highly beamed, periodic pulses on timescales of a few hours \citep{2015Natur.523..568H}. UCDs also exhibit the rapid, frequency-dependent sweeping characteristic of dynamic plasma processes.
    \item Infrared Spectral Energy Distribution (SED): Because UCDs have extremely low effective temperatures ($T_{\mathrm{eff}} \lesssim 2500$~K), their thermal emission peaks firmly in the near- and mid-infrared \citep{2005ARA&A..43..195K}. Cross-matching a candidate BRaT with infrared catalogues (e.g., 2MASS or WISE) will readily identify a UCD as a bright, exceptionally red point source with an SED deeply sculpted by molecular absorption bands (such as water and methane) \citep{2001RvMP...73..719B}. 
    \item Natural Emission: Since UCDs provide an astrophysical explanation for compact gigahertz radio emission, associating a BRaT candidate with a UCD would effectively disqualify it as an unambiguous technosignature. The exception to this is if the emission exhibits a clear, measurable astrometric offset from the dwarf star, indicating a distinct planetary origin.
\end{enumerate}

\subsubsection{Protoplanetary Disks}

Protoplanetary discs form as a consequence of angular momentum conservation during the star formation process and provide natal material for planetary systems \citep{1987ARA&A..25...23S}. Their faint observational signatures at centimetre wavelengths are different from the expected signals of an advanced technological civilisation. They can be excluded as broadband SETI confounders based on the following physical and observational characteristics:

\begin{enumerate}[label=(\roman*), leftmargin=*]

    \item VLBI Morphology: Protoplanetary discs are physically extended objects, with radii reaching up to several hundred astronomical units (AU). Simulated deep radio observations of a typical disc at $125$~pc indicate that even after $1000$ hours of integration, the peak intensity at $11$~GHz is only $\sim 4~\mu$Jy~beam$^{-1}$, corresponding to a brightness temperature of just $T_b \approx 30$~K \citep{2015aska.confE.115H}. Because they are diffuse on milliarcsecond scales, global VLBI arrays will be heavily resolved, distinguishing them clearly from the point-like BRaT. 
    \item Spectral Shape: The radio continuum of a protoplanetary disc in the $1$--$15$~GHz range is intrinsically very faint, originating primarily from the thermal emission of large (centimetre-sized) dust grains and ionised stellar winds \citep{2015aska.confE.117T}. It lacks the engineered spectral non-uniformity we expect from BRaTs. 
\end{enumerate}

\subsubsection{X-ray Binaries (XRBs) and Microquasars}

Galactic X-ray binaries, comprising a compact object (a black hole or neutron star) accreting matter from a stellar companion, frequently launch powerful relativistic radio jets. In their characteristic `hard' accretion state, these steady, compact jets produce flat or slightly inverted radio continuum spectra ($S_\nu \propto \nu^{\sim 0}$) across the centimetre regime \citep{2006csxs.book..381F}. Because they are local Galactic sources, they exhibit proper motion, and their unresolved cores yield high brightness temperatures.

They can be robustly differentiated from technosignatures through the following diagnostics:
\begin{enumerate}[label=(\roman*), leftmargin=*]

    \item High-Energy Counterparts: As their name implies, the accretion processes in XRBs produce extremely luminous X-ray emission \citep{2006ARA&A..44...49R}. A cross-match with X-ray catalogues (e.g., from \textit{Chandra} or \textit{XMM-Newton}) will readily reveal the high-energy signature of an accretion disc, which is entirely absent from a planetary technosignature model.
    \item VLBI Morphology: While the core of an XRB is highly compact, deep VLBI observations frequently resolve the extended, bi-polar structure of the relativistic jets \citep{2001MNRAS.327.1273S}, breaking the point-source degeneracy. The radio source is many orders of magnitude more luminous that a Kardashev Type I BRaT. 
    \item Spectral Shape: The flat radio spectrum of an XRB jet is a natural consequence of partially self-absorbed synchrotron emission along a conical geometry. It lacks the engineered spectral non-uniformity expected from BRaTs. 
\end{enumerate}

\subsubsection{Pulsars and Neutron Stars}

Historically, the highly periodic signals of radio pulsars provided the first major astrophysical confounders for SETI, initially designated as `Little Green Men' (LGM-1) \citep{1968Natur.217..709H}. In recent years, they have been detected in deep radio continuum surveys as relatively faint sources \citep{2016ApJ...829..119F}, and are therefore a potential confounding source. Despite their extreme compactness, several discriminants separate them from BRaTs: 

\begin{enumerate}[label=(\roman*), leftmargin=*]

    \item Spectral Shape: Pulsar emission is driven by coherent processes in the neutron star magnetosphere, universally producing very steep power-law spectra (typically $S_\nu \propto \nu^{-1.8}$) \citep{2013MNRAS.431.1352B}. Consequently, they are overwhelmingly bright at low frequencies ($\sim 100$~MHz) but rapidly fade into the GHz regime, lacking the complex spectral non-uniformity expected of a BRaT.
    \item Temporal Behaviour: Pulsars exhibit microsecond to millisecond pulse widths with extremely rigid periodicities ranging from milliseconds to seconds \citep{2005AJ....129.1993M}. This rapid, clock-like pulsation is fundamentally different from the slow, phase-stable diurnal modulation ($\sim$ hours to days) expected from a rotating planetary biosphere.
    \item Propagation Effects: As true point sources, pulsars undergo strong diffractive and refractive interstellar scintillation (ISS), producing characteristic intensity modulations across time and frequency \citep{1990ARA&A..28..561R}. While a highly compact technosignature will also undergo ISS, the inherently pulsed emission of a neutron star allows these propagation effects to be uniquely isolated and quantified. Specifically, pulsars exhibit measurable dispersion measures (DM), Faraday rotation, and temporal scattering tails that scale predictably with distance through the interstellar medium \citep{2012hpa..book.....L}. BRaTs, presenting continuous or slowly varying broadband continuum, would not yield these measurable time-of-arrival delays across frequency bands.
\end{enumerate}


\subsection{Terrestrial Radio Frequency Interference } 

Finally, a ubiquitous confounder in any radio SETI programme is terrestrial Radio Frequency Interference (RFI) generated by local telecommunications, radar, and satellite infrastructure. Interferometric arrays inherently suppress terrestrial RFI through fringe-rate modulation and delay-rate filtering \citep{2012AJ....144...38R, 2018arXiv181007235G, wandia_interferometric_2023}. Signals originating from the Earth's surface or low-Earth orbit lack the sidereal motion of celestial sources; consequently, they fail to maintain phase coherence across long baselines and rapidly decorrelate during integration. As a result, RFI that has not been removed by standard flagging processes typically contributes to the general noise floor across the synthesised field of view, rather than creating discrete artefacts in the image plane. Consequently, although RFI can elevate the noise floor and reduce survey sensitivity, it is unlikely to reproduce the compact morphology, phase stability, and persistency expected of genuine BRaT candidates.

\section{Discussion}

While the traditional SETI focus on ultra-narrowband signals remains a highly practical strategy for detecting deliberate extraterrestrial beacons, it relies on the assumption that advanced civilisations actively desire to broadcast their presence. If, instead, extraterrestrial societies are largely passive, minding their own business, searching for aggregate broadband leakage becomes essential. 


Detecting BRaTs requires a conceptual maturation of SETI, shifting the paradigm from the search for deliberate, highly conspicuous narrowband beacons to the identification of structured, planetary-scale noise. As outlined in Section 3, while natural astrophysical sources can generate faint, compact radio emission, they exhibit systematic and measurable differences from the expected properties of BRaTs. 
Distinguishing between these natural sources and BRaTs requires a nuanced approach, integrating diverse multi-wavelength observations across the electromagnetic spectrum. In this section, we propose a comprehensive diagnostic framework in which this discrimination effort is cast. 

\subsection{A Multi-Parameter Diagnostic Framework}
\label{framework}

No single observational metric is sufficient to decisively identify a broadband technosignature. 
Robust identification instead relies on the convergence of multiple independent diagnostics across a high-dimensional parameter space, breaking the physical degeneracies of individual observables. We summarise this discrimination framework in Table~\ref{tab:diagnostics}.

\subsection{Extending the Framework: Kardashev Type II Civilisations}

A Kardashev Type~II civilisation \citep{1964SvA.....8..217K} harnesses a substantial fraction of the total energy output of its host star ($P_{\mathrm{tot}} \sim 10^{26}$~W). Direct extrapolation of broadband radio leakage using the linear scaling adopted in Section~\ref{K1} is unlikely to be physically meaningful on this scale. Even if only a minute fraction of the available power were emitted or lost in the form of centimetre-wavelength radiation, the resulting emission would manifest as a bright and persistent broadband radio continuum source. 

A fundamental distinction of a Type~II BRaT lies in the spatial scale of its emitting region. For example, a Type II civilisation in the form of a distributed energy-harvesting network such as a Dyson swarm \citep{1960Sci...131.1667D} would span dimensions of the order of $\sim 1$ AU. At distances of $10$--$100$~pc, this corresponds to angular sizes of $\sim 10$--$100$~mas. Consequently, the technosignature would be partially or fully resolved by global VLBI arrays at centimetre wavelengths.

While a spatially resolved Type~II BRaT might initially mimic the extended radio continuum emission of a star-forming galaxy, it can be definitively isolated using our multidimensional framework. Indeed, the systematic application of this framework to deep radio survey data can place stringent constraints on or entirely rule out the presence of Type II BRaTs within the local Milky Way. Furthermore, we note that complementary optical photometry of such systems would likely reveal highly anomalous light curves, reflecting the transit signatures and engineered geometry of orbiting megastructures.


\subsection{The Ionospheric Barrier}

An important physical factor governing the detectability of planetary-scale radio emission is the ionospheric cut-off. The propagation of electromagnetic waves through a planetary atmosphere is fundamentally constrained by the local plasma frequency, $\nu_p$. A radio signal originating from the planetary surface or low-altitude orbit can only escape into the interstellar medium if its frequency strictly exceeds this critical threshold ($\nu > \nu_p$). The plasma frequency is determined by the maximum electron number density, $n_e$, within the ionosphere \citep{1986rpa..book.....R}:

\begin{equation}
\nu_p = \frac{1}{2\pi} \sqrt{\frac{n_e e^2}{m_e \epsilon_0}} \approx 8.98 \sqrt{n_e} \,\, \mathrm{Hz}
\end{equation}

where $e$ is the elementary charge, $m_e$ is the electron mass, $\epsilon_0$ is the vacuum permittivity, and $n_e$ is expressed in $\mathrm{m}^{-3}$. 

On Earth, the dayside ionosphere typically reaches peak electron densities of $n_e \sim 10^{12}~\mathrm{m}^{-3}$ \citep{2009eipp.book.....K}, resulting in a characteristic radio escape cut-off of $\nu_p \approx 10$~MHz. However, ionospheric electron densities are driven primarily by photoionisation from the incident extreme-ultraviolet (EUV) and X-ray flux of the host star. Exoplanets within the habitable zones of active M-dwarfs experience severe and sustained EUV fluxes \citep{2016ApJ...820...89F, 2019ApJ...871..235P}. Extrapolating from atmospheric models of highly irradiated exoplanets (e.g., hot Jupiters), it is plausible that peak electron densities can exceed those observed on Earth by several oders of magnitude e.g. $n_e \gtrsim 10^{14}~\mathrm{m}^{-3}$ \citep{2014ApJ...796...16K}. Although further characterisation of terrestrial exoplanet ionospheres is required, such high electron densities would undoubtedly push the local radio blackout limit into the hundreds of MHz. Under these conditions, low-frequency technological leakage would be entirely trapped beneath the plasma shield via internal reflection. Searching for BRaTs at $\gtrsim 1$~GHz mitigates this risk.

We note that the presence of a sharp, low-frequency ionospheric cut-off within a broadband spectrum could itself act as a powerful diagnostic tool. If this spectral cut-off exhibits phase-stable diurnal variation, driven by the rotating planet's varying exposure to stellar photoionisation, it would strongly support the artificial, planetary-surface origin of the radio emission.

\subsection{AI-Driven Anomaly Detection}

As described in Section~\ref{framework}, robust identification of broadband technosignatures requires simultaneous evaluation of the multidimensional diagnostics detailed in Table~\ref{tab:diagnostics}. Applying this complex analysis to the hundreds of millions of compact radio sources anticipated from next-generation wide-field surveys necessitates the deployment of Artificial Intelligence (AI) and machine learning (ML) pipelines specifically tailored for high-dimensional anomaly detection. By ingesting vast, multi-wavelength catalogues that incorporate astrometry, polarisation fractions, spectral topology, and temporal variability metrics, unsupervised algorithms can construct a highly accurate statistical representation of the natural astrophysical background, enabling them to separate structured artificial anomalies from standard unstructured noise. This will enable the identification of isolated outliers using the appropriate combination of physical and geometric properties expected of the planetary-scale technological aggregate we envisage for BRaTs. Implementing this will be challenging, as automated pipelines must be designed to handle the inherent heterogeneity of astronomical data. However, it is highly improbable that every source in a wide-field survey will immediately possess the complete suite of high-resolution observables required by the diagnostic matrix. By flagging highly anomalous continuum sources in lower-resolution surveys, these algorithms can automatically trigger the deep, targeted VLBI follow-up required to confirm a candidate's astrometric and temporal technosignature credentials.

\section{Conclusions}

Traditional SETI strategies have emphasised the search for narrowband radio (and optical) signals because they are readily distinguishable from natural astrophysical emission. In addition, deliberately engineered narrowband beacons remain an energetically efficient and physically well-motivated signalling strategy over interstellar distances (e.g. \citet{2025AJ....169..118S}. However, as our own technological trajectory demonstrates, advanced communication infrastructures may naturally evolve toward distributed, broadband systems.  To avoid systematically misclassifying such aggregate leakage as unstructured cosmic noise, we must expand our search parameters to identify broadband radio technosignatures (BRaTs). This approach is complementary to conventional narrowband SETI methodologies. 

We have demonstrated that a tiered hierarchical framework, utilising deep wide-field surveys by the SKA and other wavelengths, followed by targeted global VLBI observations, provides the unparalleled sensitivity, spatial filtering, astrometric precision, and extreme brightness-temperature sensitivity required to isolate these signals.

A fundamental observational advantage of searching for BRaTs is their intrinsic insensitivity to Doppler drift. This permits arbitrarily long integrations, enabling the concept of ``SETI Deep Fields.'' Such prolonged campaigns can improve sensitivity by over an order of magnitude, bringing faint Kardashev Type~I planetary leakage within the detectable volume ($\lesssim 100$~pc) of next-generation arrays. Furthermore, we have suggested  that observing at frequencies $\gtrsim 1$~GHz bypasses the dense ionospheric plasma barriers expected around planets orbiting active M-dwarf stars, ensuring that aggregate leakage remains visible to interstellar observers.

To distinguish candidate BRaTs from the dense population of discrete cosmic confounders such as highly compact AGN, distant starforming galaxies, pulsars, and active local stars or magnetised exoplanets, we have presented a comprehensive multi-parameter diagnostic framework. Genuine BRaTs will be identified not by a single metric, but through a distinct convergence of physical observables: extreme spatial compactness yielding $T_b \gtrsim 10^6$~K, negligible polarisation, engineered spectral non-uniformity, phase-stable diurnal modulation superimposed upon interstellar scintillation, and sub-milliarcsecond astrometric co-motion with an offset local stellar host. We note that this framework can also be extended to Kardashev Type~II civilisations, which may appear as spatially resolved, AU-scale technological aggregates. 

The transition to wide-field broadband SETI necessitates a fundamental evolution in data analysis. The sheer volume and dimensionality of next-generation astronomical surveys will require the deployment of unsupervised, AI-driven anomaly-detection pipelines. By modelling statistical representations of natural astrophysical populations, these algorithms can perform hierarchical triage, flagging multidimensional outliers to automatically trigger targeted VLBI follow-up. 

Crucially, the implementation of this framework need not wait for the full deployment of the SKA. Current deep radio continuum surveys conducted by SKA precursor facilities offer an immediate opportunity to refine and deploy these diagnostic pipelines. By interrogating existing multi-wavelength datasets, we can actively test this paradigm today, allowing us to place the first direct observational constraints on the prevalence and luminosity of BRaTs within the local Galactic environment.

Broadening our search strategies to include the unintentional, aggregate radio footprints of advanced civilisations represents a necessary maturation of the field. By combining the resolving power of global interferometric networks with rigorous physical diagnostics and machine learning, we significantly enhance our capacity to detect the complex, structured noise of technologically advanced biospheres hiding within the cosmic background.

\section*{Acknowledgements}

The author thanks the anonymous reviewer for their careful reading and constructive comments, which improved this manuscript.

\section*{Data Availability}

No new observational data were generated or analysed in support of this research. The theoretical framework and sensitivity estimates presented in this article were derived using publicly available tools, including the European VLBI Network (EVN) Sensitivity Calculator.



\bibliographystyle{mnras}
\bibliography{References.bib} 




\begin{landscape}

\begin{table}
\centering
\caption{Diagnostic matrix for distinguishing broadband radio technosignatures (BRaTs) from discrete astrophysical confounders in the 1--10 GHz regime.}
\label{tab:diagnostics}
\renewcommand{\arraystretch}{1.3}
\begin{tabular}{>{\raggedright\arraybackslash}p{2.8cm} p{2.2cm} p{1.5cm} p{2.5cm} p{1.8cm} p{3cm} p{2.8cm}}
\hline
\textbf{Source Type} & \textbf{Morphology (VLBI)} & \textbf{$T_b$ Limit} & \textbf{Spectral Shape} & \textbf{Polarisation} & \textbf{Temporal Behaviour} & \textbf{Astrometry \& Kinematics} \tabularnewline
\hline
\textbf{BRaTs (Type I)} & Unresolved ($\sim\mu$as) & $>10^6$~K & Complex, non-uniform, atmospheric notches & Negligible / Weak & Phase-stable diurnal modulation + local ISS & Local ($<100$~pc), co-moving, astrometric offset from star \tabularnewline
\textbf{Active Galactic Nuclei (AGN)} & Unresolved / Core-jet & $>10^6$~K & Smooth power-law & Variable & Stochastic flares + Galactic ISS (IDV) & Cosmological, stationary background \tabularnewline
\textbf{Star-forming Galaxies (SFGs)} & Extended ($\sim$arcsec) & $<10^6$~K & Smooth power-law & Weak / Unpolarised & Steady & Cosmological, stationary background \tabularnewline
\textbf{Magnetised Exoplanets} & Unresolved & High & Sharp high-frequency cut-off ($\lesssim$ MHz) & Highly circular & Bursty, rotationally modulated & Local, co-moving, astrometric offset from star \tabularnewline
\textbf{Active Stars (e.g., M-dwarfs)} & Unresolved & High & Peaked / Smooth (incoherent) & Strong circular (coherent bursts) & Stochastic flaring & Local, co-moving, spatially coincident with star \tabularnewline
\textbf{Ultracool Dwarfs (Brown Dwarfs)} & Unresolved & High & Sharp high-frequency cut-off ($\sim$ GHz) & 100\% Circular & Beamed, rotationally modulated pulses & Local, co-moving, coincident with IR source \tabularnewline
\textbf{X-ray Binaries (XRBs) \& Microquasars} & Compact core / Resolved jets & High & Flat / Inverted ($S_\nu \propto \nu^{\sim 0}$) & Linear / Variable & Steady (hard state) / Stochastic flares & Galactic (kpc scales), negligible proper motion at $\mu$Jy levels, bright X-rays \tabularnewline
\textbf{Pulsars \& Neutron Stars} & Unresolved & $>10^{10}$~K & Steep power-law & Strong Linear / Circular & Millisecond-second periodic pulses & Galactic, high independent proper motion, highly dispersed \tabularnewline
\textbf{Protoplanetary Discs} & Extended ($\sim$100~AU) & $\sim30$~K & Thermal dust / free-free emission & Unpolarised & Steady & Local, co-moving (young stellar systems) \tabularnewline
\hline
\end{tabular}
\end{table}

\end{landscape}

\bsp	
\label{lastpage}
\end{document}